\documentclass[aps,reprint,superscriptaddress,longbibliography,floatfix]{revtex4-2}
\usepackage{amsmath,amssymb,bm,graphicx,xcolor}
\usepackage[colorlinks=true,citecolor=blue,linkcolor=blue,urlcolor=blue]{hyperref}
\newcommand{\av}[1]{\left\langle #1\right\rangle}
\newcommand{\W}{\mathcal W}
\input{glyphtounicode}
\begin{document}

\title{Witnessing Gaussian Coexistence and Correlation-Ellipse Tomography}
\author{Wanting Hou}
\thanks{These authors contribute equally to this work.}
\author{Wei Liu}
\thanks{These authors contribute equally to this work.}
\affiliation{Department of Physics, Faculty of Arts and Sciences, Beijing Normal University, Zhuhai 519087, China}
\author{Hong-Chao Liu}
\affiliation{Institute of Applied Physics and Materials Engineering, University of Macau, Macao SAR, 999078, China}
\author{Jun Xiong}
\affiliation{School of Physics and Astronomy, Applied Optics Beijing Area Major Laboratory, Beijing Normal University, Beijing 100875, China}
\author{Zhiyuan Ye}
\email{yezy@bnu.edu.cn}
\affiliation{Department of Physics, Faculty of Arts and Sciences, Beijing Normal University, Zhuhai 519087, China}

\begin{abstract}
Gaussian light is specified by second-order field moments, yet their hidden correlation geometry can demand higher-order intensity measurements. We establish a measurement hierarchy for ``Gaussian coexistence'', i.e., the simultaneous presence of cross covariance and cross pseudocovariance. For two zero-mean Gaussian modes with circular marginals, all unperturbed intensity moments of second or third order are blind to coexistence. In contrast, a balanced fourth-order residual provides a necessary and sufficient witness.
Their identities, however, remain fundamentally ambiguous: recovering the two correlation strengths does not reveal which belongs to the covariance and which to the pseudocovariance.
Unexpectedly, adding thermal fluctuations resolves this ambiguity: an independent, correlated thermal probe selectively modifies the covariance while leaving the pseudocovariance unchanged, allowing the two correlation strengths to be individually identified.
Moreover, a coherent reference in only one arm then makes their relative phase accessible through third-order intensity correlation. Together, these measurements reconstruct a correlation ellipse and its Poincar\'e-sphere coordinates, determining the relative channel strengths and the ellipse orientation. The framework separates what intensity detection can witness, what thermal probing can identify, and what a single phase reference can orient, turning inaccessible Gaussian correlation structure into a quantitatively reconstructible geometry.
\end{abstract}
\maketitle

\subsection*{Introduction}
Gaussian statistics compress an optical field's fluctuating complexity into its mean and second-order field moments~\cite{Goodman2015}.
Hanbury Brown-Twiss intensity interferometry exploits this structure to infer coherence without directly measuring the optical phase~\cite{HanburyBrown1956,Ferreira2020}. For a complex field, however, covariance and pseudocovariance are independent parts of the second-order description~\cite{NeeserMassey1993,Picinbono1996,SchreierScharf2003}. Their distinct optical responses underlie Gaussian coherence theory and can support complementary image-encoding channels~\cite{ErkmenShapiro2008,verniere2024hiding,ye2024hanbury,Hou2026}.
For zero-mean, stationary, jointly Gaussian fields $E_1,E_2$ with circular marginals ($\av{E_i^2}=0$), define $I_i=|E_i|^2$, $n_i=\av{I_i}>0$, $\Gamma=\av{E_1E_2^*}$, and $C=\av{E_1E_2}$. Gaussian moment factorization gives the relation for the joint noncircular case~\cite{ErkmenShapiro2008,lemieux1999investigating}
\begin{align}
 G^{(2)}&\equiv\av{I_1I_2}=\av{E_1E_1^*E_2E_2^*}=n_1n_2+|\Gamma|^2+|C|^2,
 \label{eq:siegert}
\end{align}
which retains both covariance and pseudocovariance contributions.
The familiar second-order intensity measurement $G^{(2)}$ sees only a sum of squared amplitudes. Here we call their simultaneous presence, \(\Gamma\neq0\) and \(C\neq0\), \emph{Gaussian coexistence}. This constitutes an experimentally accessible yet often overlooked structure within Gaussian correlations, beyond the requirement \(C\neq0\) for joint noncircularity~\cite{picinbono1994circularity}. Counterintuitively, two coexisting correlation channels can produce weaker normalized intensity bunching $g^{(2)}$ than a fully correlated single channel~\cite{Hou2026}. Reduced bunching alone, however, cannot establish $C$-$\Gamma$ coexistence: a single channel with incomplete mutual coherence can produce the same degree of $g^{(2)}$~\cite{lachs1965theoretical,singer2013hanbury,kondakci2016sub,han2023tailoring}. The experimental challenge is therefore to distinguish these physically different structures: what measurements can unambiguously establish that both channels are present and resolve their individual contributions? More subtly, \(\Gamma\) and \(C\) themselves are complex quantities, prompting us to ask whether their relative phase can be measured and what physical information they convey.

Higher-order correlations have already furnished powerful diagnostics, from photon bunching and blockade to quantum non-Gaussianity and many-body fluctuations~\cite{Assmann2009,avenhaus2010accessing,dall2013ideal,pearce2015precision,Hamsen2017,Schweigler2017,liang2018observation,schweigler2021decay,Xia2025,Hotter2026}. In particular, the use of $g^{(2)}$ and $g^{(3)}$ to exclude mixtures of Gaussian quantum states~\cite{Hotter2026} motivates asking which combinations of measured correlations resolve structure {within} a classical Gaussian model.
Also, the reference-assisted reconstruction and photon-statistical characterization of Gaussian states are well established~\cite{ArkhipovPerina2016,RuppertUsenkoFilip2016,Thekkadath2022Displacement,HeinzelSondenheimer2026}. Here we resolve a more specific question: which features of covariance-pseudocovariance coexistence are accessible from the intensity statistics, and which additional probing resources are required to identify and reconstruct its correlation geometry.
We establish a measurement hierarchy: fourth-order moment witnesses $C$-$\Gamma$ coexistence, thermal probing assigns its channel strengths, and a single-arm reference recovers the relative phase via third-order intensity correlation.
Together, the measured channel strengths and relative phase reconstruct a \textit{correlation ellipse}, whose geometry follows from the superposition of two counter-rotating complex components~\cite{SchreierEllipse2008}. We term this reconstruction \textit{correlation-ellipse tomography}: it determines the shape and orientation of the conditional-mean trajectory linking the two Gaussian fields.

\subsection*{Correlation ellipse and sphere}
\begin{figure}[htbp]
 \centering
 \includegraphics[width=\columnwidth]{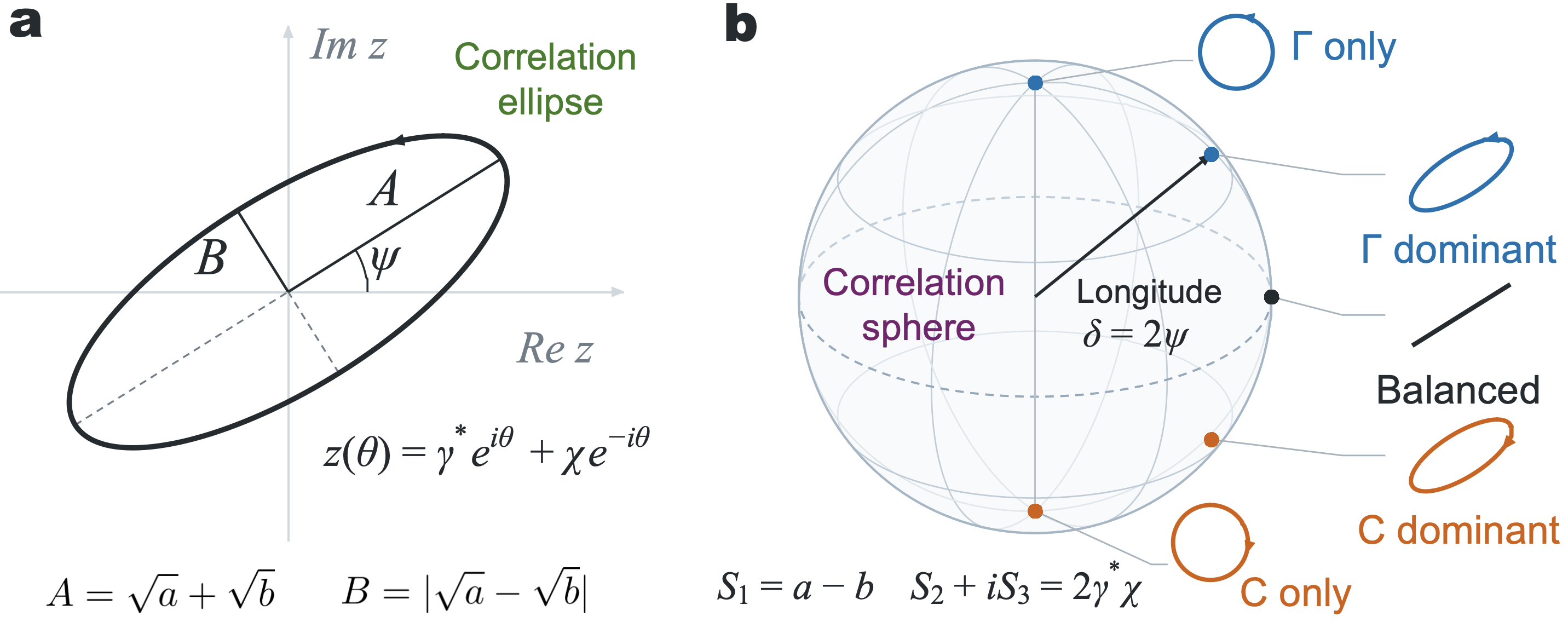}
 \caption{Conceptual illustration of correlation geometry. The counter-rotating covariance and pseudocovariance contributions generate the conditional-mean trajectory in Eq.~\eqref{eq:ellipse}. Their balance and relative phase specify the ellipse and the normalized correlation sphere. The poles are single-channel circles with opposite traversal senses; equal strengths lie on the equator and give a line. %
 }
 \label{fig:ellipse}
\end{figure}

Previous coherence Poincar\'e spheres describe vector-field coherence without incorporating pseudocovariance~\cite{Laatikainen2021, Laatikainen2022}.
To give these two channels an intuitive geometric interpretation, we construct an ellipse from the conditional mean of one field as the phase of the other is scanned, and represent its geometry on a Poincar\'e sphere. This construction translates the channel strengths and relative phase into the ellipse’s shape and orientation, providing geometric coordinates for the correlation tomography developed below.
Throughout, %
we set $\gamma=\Gamma/\sqrt{n_1n_2}$, $\chi=C/\sqrt{n_1n_2}$, $a=|\gamma|^2$, $b=|\chi|^2$, and $p=a+b$ represents the overall correlation strength.
We consider the conditional mean of $E_2$ as the phase \(\theta\) of $E_1$ varies at a fixed conditioning amplitude \(\sqrt{n_1}\). This defines a trajectory in the complex-amplitude plane:
\begin{equation}
 z(\theta)\equiv\frac{\mathbb E[E_2\mid E_1=\sqrt{n_1}\mathrm{e}^{\mathrm{i}\theta}]}{\sqrt{n_2}}
 =\gamma^*\mathrm{e}^{ \mathrm{i}\theta}+\chi \mathrm{e}^{- \mathrm{i}\theta}.
 \label{eq:ellipse}
\end{equation}
The two cross moments generate counter-rotating contributions, analogous to the circular components of a vector field. Their sum traces a \emph{conditional-mean correlation ellipse} as shown in Fig.~\ref{fig:ellipse}(a). Its semiaxes are $A=\sqrt a+\sqrt b$ and $B=|\sqrt a-\sqrt b|$, its orientation is $\psi=\arg(\gamma^*\chi)/2$, and its traversal sense is set by $a-b$.
At fixed channel strengths, a change \(\Delta\delta\) in \(\delta=\arg(\Gamma^*C)\) rotates the conditional-mean ellipse by \(\Delta\delta/2\) while leaving all original intensity moments unchanged.
This is a two-field regression geometry: the ellipse describes how the conditional mean of $E_2$ varies with the phase of $E_1$, while each field individually retains circular Gaussian statistics.

Introduce Stokes-like correlation coordinates
\begin{equation}
 S_0=p,\qquad S_1+\mathrm{i}S_2=2\gamma^*\chi,\qquad S_3=a-b,
 \label{eq:stokes}
\end{equation}
which satisfy $S_1^2+S_2^2+S_3^2=S_0^2$. For $p>0$, $\bm s=(S_1,S_2,S_3)/p$ lies on a unit Poincar\'e sphere as shown in Fig.~\ref{fig:ellipse}(b): longitude specifies $2\psi$, and the (northern or southern) hemisphere identifies the dominant channel and traversal sense. Note that the analogy does not introduce a new physical polarization but depicts the geometric relationship between two complex cross moments.
Nor does a circular trajectory imply joint circular statistics: a $C$-only Gaussian correlation is jointly noncircular but occupies the southern pole. Equal amplitudes give a degenerate line, still a $C$-$\Gamma$ coexistence correlation.
The regression derivation and comparison with polarization ellipses are given in Sec.~S2 of the Supplemental Material.

\subsection*{Fourth-order witness of coexistence}
\begin{figure}[htbp]
 \centering
 \includegraphics[width=\columnwidth]{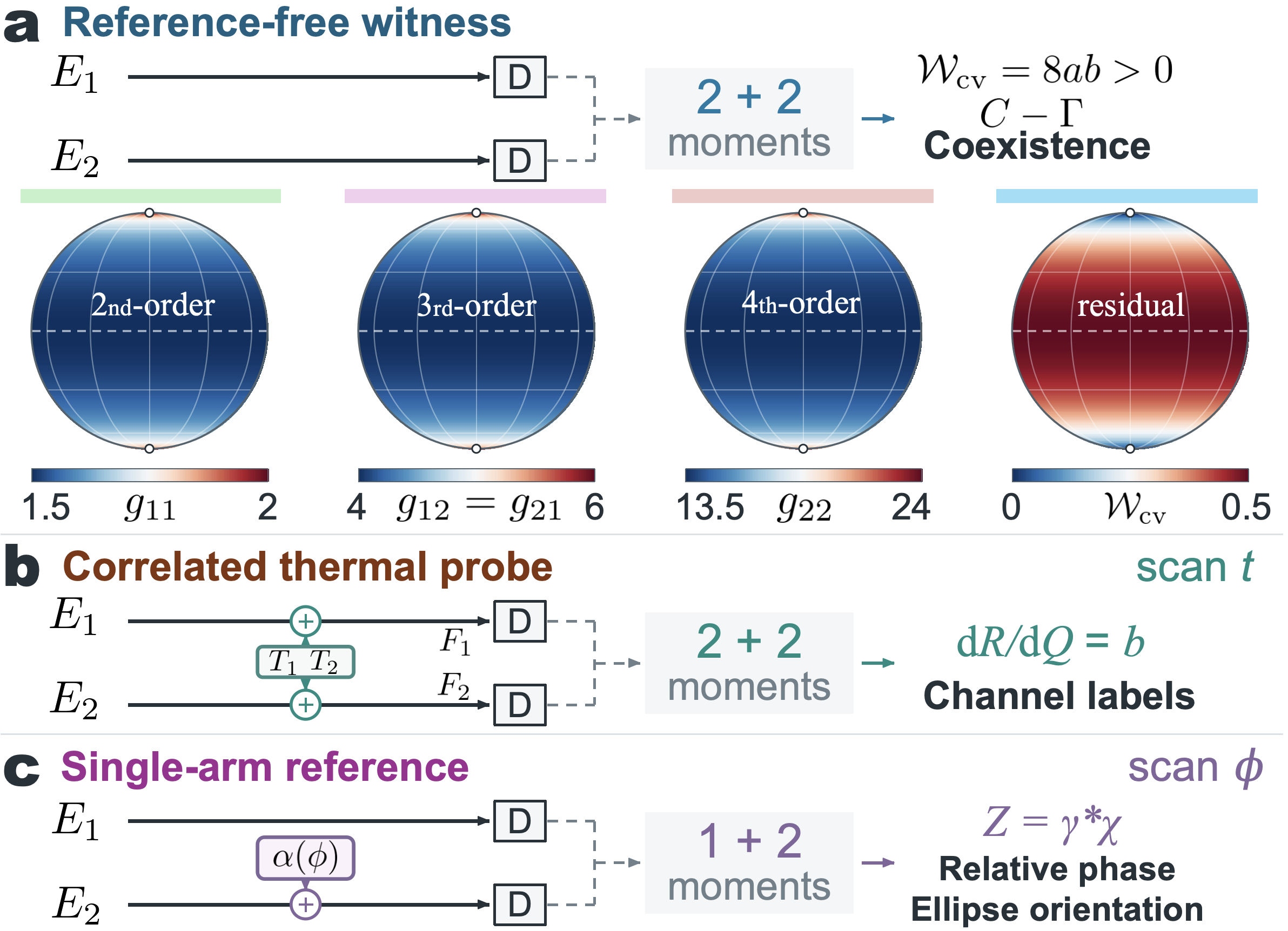}
 \caption{Measurement hierarchy. (a) Reference-free moments yield $\W_{\rm cv}$; sphere maps below show $g_{11}$, $g_{12}$ ($g_{21}$), $g_{22}$, and $\W_{\rm cv}$ for the ideal mixture $\sqrt a+\sqrt b=1$. North/south poles denote $\Gamma$/$C$-only Gaussian correlations; dashed equators mark equal strengths. (b) An independent, correlated proper thermal probe labels $b$ through the $R$--$Q$ slope. (c) A single-arm reference recovers $Z=\gamma^*\chi$ by phase cycling.
 }
 \label{fig:witness}
\end{figure}

Figure~\ref{fig:witness} summarizes the three measurement configurations.
Let $g_{mn}=\av{I_1^mI_2^n}/(n_1^m n_2^n)$, where $m+n$ is the total intensity order.
An ideal circular Gaussian field has an exponentially distributed intensity, giving \(g_{m0}=m!\) and \(g_{0n}=n!\).
Gaussian factorization yields $g_{11}=1+p$, $g_{21}=g_{12}=2+4p$, and
\begin{align}
 g_{22}=4+16p+4p^2+8ab.
 \label{eq:moments}
\end{align}
All moments through third order depend on the same $p$, %
whereas balanced fourth-order moment in Eq.~(\ref{eq:moments}) supplies the missing product:
\begin{equation}
 {\W\equiv g_{22}-4g_{11}^{\,2}-8g_{11}+8=8ab.}
 \label{eq:witness}
\end{equation}
Within the stipulated Gaussian class, $\W>0$ is \emph{necessary and sufficient} for $C$-$\Gamma$ coexistence.
Only the balanced $2+2$ measurement matters, while the unbalanced fourth-order moments $g_{31}=g_{13}=6+18p$ remain blind to the product.
Sections~S3.B and S3.C of the Supplemental Material give the proof.
The sphere maps in Fig.~\ref{fig:witness}(a) use the ideal mixture $\sqrt a+\sqrt b=1$, with $p=[1+\sqrt{1-s_3^2}]^{-1}$. Coexistence lowers $g_{11}$ from 2 at either pole to 1.5 at equal mixing (the equator). However, $g_{11}<2$ is necessary but not sufficient for $C$-$\Gamma$ coexistence and can be found in many circumstances~\cite{lachs1965theoretical,singer2013hanbury,kondakci2016sub,han2023tailoring}.

Positivity, $\sqrt a+\sqrt b\le1$, imposes the tight bound $0\le\W\le2\min\{p^2,(1-p)^2\}\le1/2$.
Based on the measurements of $p$ and $\W$, one determines
\begin{equation}
 \{a,b\}=\left\{{p/2\pm\sqrt{p^2/4-\W/8}}\right\},
 \label{eq:roots}
\end{equation}
but leave the roots unlabeled. Indeed, $E_2\mapsto E_2^*$ exchanges $\Gamma$ and $C$ while preserving every intensity sample. No order of the original intensity statistics can remove this ambiguity, and arbitrary local phase rotations also leave them unchanged. On the correlation sphere, $p$ and $\W$ fix the scale and absolute latitude, but neither the hemisphere nor longitude.

Third-order moments, although unable to witness coexistence alone, can reduce the sampling noise of the fourth-order test.
By estimating all moments from the same synchronized record, one forms the Gaussian closure residuals $\Delta_3=g_{21}+g_{12}-8 g_{11}+4$ and $\Delta_2= g_{20}+ g_{02}-4$. Their population values vanish, so they provide control variates:
\begin{equation}
 \W_{\rm cv}=\W-4\Delta_3+2\Delta_2
 \ \xrightarrow \ 8ab.
 \label{eq:controlvariate}
\end{equation}
This estimator retains the same population witness without additional optical settings.
Its construction and sampling uncertainty are derived in Secs.~S3.F--S3.G.

\subsection*{Thermal probing and channel identification}
To assign the two roots in Eq.~\eqref{eq:roots} to their physical channels, we use thermal fluctuations themselves as a probe of the random light. The key is to introduce randomness with a controlled correlation structure. As shown in Fig.~\ref{fig:witness}(b), we superpose the signal with a jointly proper Gaussian thermal-light pair $T_1,T_2$, which is internally correlated but statistically independent of $E_1,E_2$. Writing $F_i(t)=E_i+\sqrt{t}\,T_i$, we define $m_i=\av{|T_i|^2}$ and $d=\av{T_1T_2^*}\ne0$, with $\av{T_iT_j}=0$ and adjustable probe-power scaling $t$. Statistical independence then gives $N_i=n_i+tm_i$, $\Gamma_t=\Gamma+td$, and $C_t=C$. Thus, the added thermal randomness selectively changes the covariance while leaving the pseudocovariance unchanged, making the two channel strengths individually identifiable without a coherent local oscillator.

Using the normalized moments and measured mean-intensity factor
\begin{equation}
 g_{mn}^{\rm th}(t)=\frac{\av{|F_1|^{2m}|F_2|^{2n}}}{N_1^mN_2^n},
 \qquad \lambda(t)=\frac{N_1N_2}{n_1n_2},
 \label{eq:probe}
\end{equation}
and defining $\W_{\rm cv}^{\rm th}$ by Eqs.~\eqref{eq:witness}\eqref{eq:controlvariate} using $g_{mn}^{\rm th}$, we now construct the dimensionless combinations
\begin{align}
 Q(t)=\lambda(t)[g_{11}^{\rm th}(t)-1], \,\,\,
 R(t)=\lambda(t)^2\W_{\rm cv}^{\rm th}(t)/8.
 \label{eq:QR}
\end{align}
The factors of $\lambda$ account for the probe-induced change in the mean intensities. The measured combinations obey
\begin{equation}
 {R=bQ-b^2,\qquad a=Q(0)-b.}
 \label{eq:line}
\end{equation}
Thus, the $R$--$Q$ slope directly identifies the previously unlabeled normalized pseudocovariance strength $b$, without requiring the complex value of $d$. Since $Q(0)=p$, the correlation weights are $w_C=b/p$ and $w_\Gamma=a/p=1-w_C$.
In geometrical terms, $S_3=Q(0)-2b$ now identifies the sphere's hemisphere.
Refer to Sec.~S4 for detailed derivations.

\subsection*{Single-reference ellipse tomography}
Thermal probing determines the amplitudes but not the ellipse orientation.
As shown in Fig.~\ref{fig:witness}(c), return to the unperturbed pair and add a coherent reference $\alpha(\phi)$ only to $E_2$: $J_2(\phi)=|E_2+\sqrt P \mathrm{e}^{\mathrm{i}\phi}|^2$, with known reference intensity $P>0$, $\rho=P/n_2$ and adjustable phase shift $\phi$. The normalized third-order intensity correlation is
\begin{equation}
 \begin{aligned}
 g_{12}^{\rm ref}(\phi)\equiv\frac{\av{I_1J_2(\phi)^2}}{n_1(n_2+P)^2}
 =\frac{  4\rho }{(1+\rho)^2} \operatorname{Re}\left[\mathrm{e}^{-2\mathrm{i}\phi}\gamma^*\chi\right]+B_0
 \end{aligned}
 \label{eq:third}
\end{equation}
with the background of $B_0=(g_{12}+4\rho g_{11}+\rho^2)/(1+\rho)^2$.
When both channels are present, the phase-dependent term oscillates as \(\cos[2\phi-\arg(\gamma^*\chi)]\), producing a fringe with a period of \(\pi\).
In turn, a nonzero fringe visibility therefore provides a reference-assisted witness of $C$-$\Gamma$ coexistence (refer to Sec.~S5.C).
This second-harmonic response arises from the quadratic dependence on the coherent reference. Its phase offset directly determines the correlation-ellipse orientation, which is defined modulo \(\pi\).
Moreover, four-step reference phases at the same $P$ cancel the phase-independent background and give
\begin{equation}
 \begin{aligned}
 &Z\equiv\frac{(1+\rho)^2}{8\rho}\bigl\{g_{12}^{\rm ref}(0)-g_{12}^{\rm ref}(\pi/2)\\
 &+\mathrm{i} \left [g_{12}^{\rm ref}(\pi/4)-g_{12}^{\rm ref}(3\pi/4) \right]\bigr\}
 =\gamma^*\chi.
 \end{aligned}
 \label{eq:phasecycle}
\end{equation}
One thus has $S_1+\mathrm{i}S_2=2Z$ and $\psi=\arg Z/2$. Together with the thermal reconstruction, these quantities specify the ellipse scale, axes, orientation, and traversal sense. A second-order cross-intensity measurement with only a single-arm reference has no phase-dependent term, so third order is the first such phase-sensitive mixed moment in this configuration.
A single-arm reference still cannot label the two strengths, because $E_1\mapsto E_1^*$ exchanges $(\Gamma,C)$ with $(C^*,\Gamma^*)$ while preserving the measured data and $Z$. Thus, thermal probing supplies precisely this missing label (see also Sec.~S5.D).
This completes the measurement hierarchy in theory: fourth-order moment witnesses coexistence, a thermal probe labels the channel strengths, and third-order moment recovers their relative phase with a single-arm reference. The correlation ellipse connects these distinct resources to the shape, orientation, and traversal sense of a two-field regression geometry.

\begin{figure}[!htbp]
 \centering
 \includegraphics[width=0.92\columnwidth]{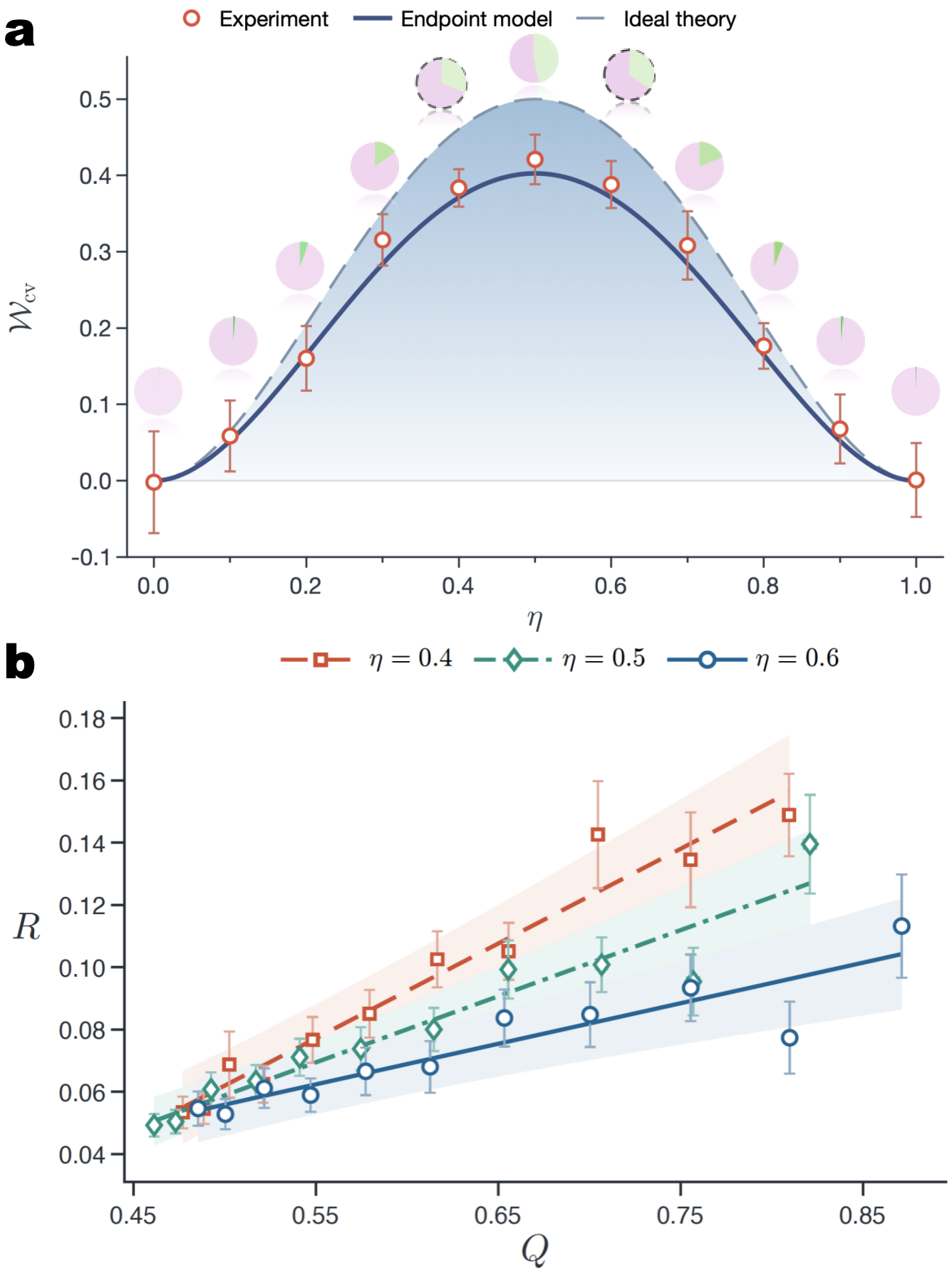}
 \caption{Fourth-order detection of coexistence and thermal identification of the two channels. (a) Measured $\W_{\rm cv}$ versus the covariance-source power fraction $\eta$.
 The solid curve (endpoint model) uses $g_{11}$ measured at $\eta=0$ and $1$ to predict $\W$ at intermediate mixtures. The dashed boundary of the shading gives the ideal prediction ($u=v=1$). Pink and green sectors show the relative weights of the two channels. (b) Thermal-probe scans for the three outlined settings in (a): symbols show $R$-$Q$ values at eleven probe-power settings; Lines are ordinary least-squares fits; shaded regions indicate approximate pointwise 95\% confidence intervals.
 }
 \label{fig:ratio}
 \label{fig:3}
\end{figure}

\subsection*{Experimental observations}
To implement this measurement hierarchy, an expanded coherent beam illuminates a spatial light modulator (SLM) displaying a sequence of hybrid random-phase patterns, producing a programmable Gaussian source with a tunable $C$-$\Gamma$ balance. Each pattern combines two statistically independent phase masks on complementary sets of equal-area annuli: (1) a centrosymmetric random-phase pattern~\cite{he2025label}, $\phi_\Gamma(\bm r)=\phi_\Gamma(-\bm r)$, generates equal Fourier-plane fields at opposite coordinates and supplies the covariance channel; (2) a binary random-phase pattern, $\phi_C(\bm r)\in\{0,\pi\}$, gives a real pupil field and hence conjugate Fourier-plane fields, supplying the pseudocovariance channel.
The annuli are randomly assigned to the two sets, with this partition held fixed as the phase patterns evolve.
An optical Fourier lens projects the SLM plane onto the far field via a 2f system, and the intensity statistical fluctuations among sets of centrosymmetric points are measured for the subsequent calculation of higher-order moments.
The relative power of the two channels is controlled by reassigning the annuli without tuning either isolated source’s intrinsic correlation. See also Sec.~S7 for more details.
The mixing fraction \(\eta\) denotes the fraction of the detected mean intensity contributed by the $\Gamma$ component in the Gaussian source and can be approximated by its assigned annular area fraction.
Let \(u\) and \(v\) denote the fixed normalized correlation strengths of the $\Gamma$-only and $C$-only source components, respectively. Their normalized second-order cross-intensity correlations are therefore \(1+u\) and \(1+v\).
For the same covariance-source power fraction $\eta$ in both detected modes, we thus have $a=u\eta^2$, $b=v(1-\eta)^2$, and $\W(\eta)=8uv\eta^2(1-\eta)^2$.

\begin{figure*}[htbp]
 \centering
 \includegraphics[width=0.96\textwidth]{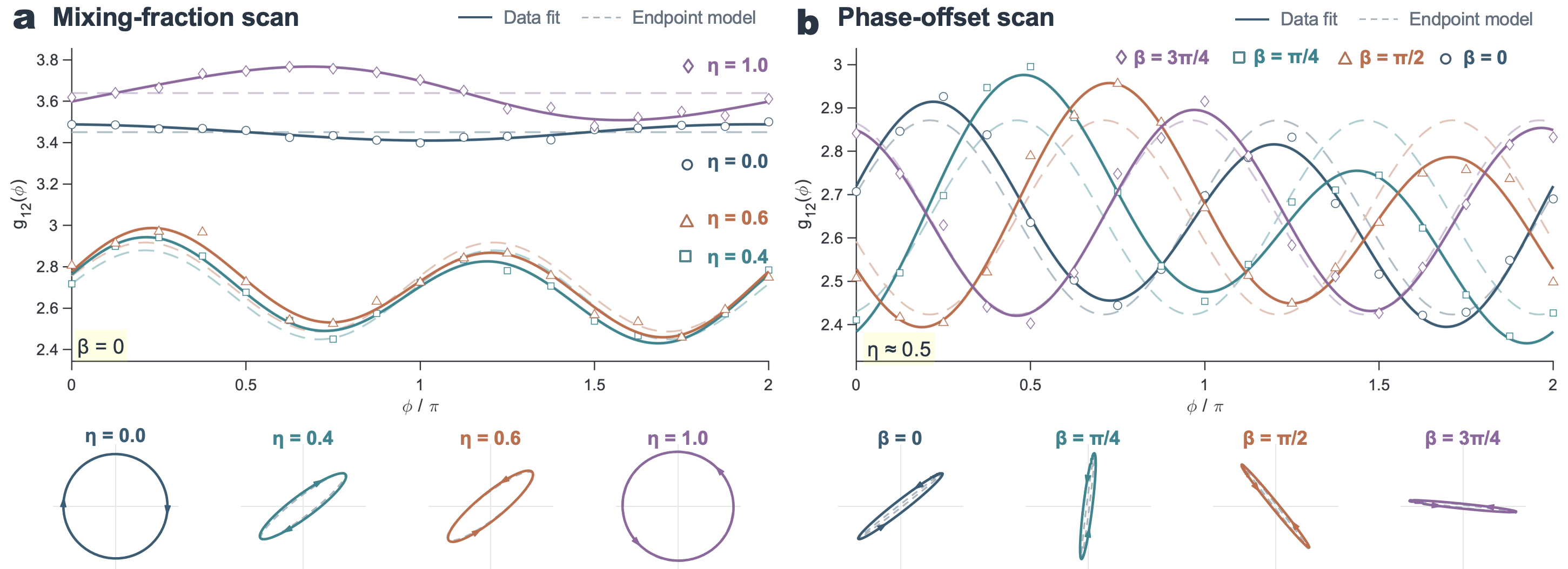}
 \caption{Single-reference reconstruction of eight correlation ellipses via $g_{12}^{\rm ref}(\phi)$. (a) Mixing-fraction scan at $\beta=0$ for $\eta=0,0.4,0.6,1$. (b) Phase-offset scan at $\eta\simeq0.5$ for $\beta=0,\pi/4,\pi/2,3\pi/4$. Solid curves are fits retaining first and second harmonics, and pale dashed curves are endpoint-model predictions. Below, solid ellipses in the complex $z$ plane are reconstructed from the fitted background and second harmonic and compared with the same model (dashed).
 }
 \label{fig:4}
\end{figure*}

For each source partition $\eta$ in the experiment, we first evaluate $\W_{\rm cv}$ separately for ten pairs of centrosymmetric points using $10^5$ synchronized intensity samples per pair, giving $10^6$ paired samples in total per run.
The measured witness is consistent with zero at both single-channel endpoints and reaches the peak of $0.421\pm0.033$ at $\eta=0.5$ [Fig.~\ref{fig:3}(a)]. The fact that the peak falls below the ideal-source peak of 0.5 is mainly due to the \(g_{11}\) of the \(\Gamma\) and \(C\)-only components (the two endpoints) being below the ideal value of 2. In the endpoint-calibrated model, we take $u=g_{11}-1=0.945$ at the covariance-only endpoint ($\eta=1$) and $v=g_{11}-1=0.850$ at the pseudocovariance-only endpoint ($\eta=0$) for each point pair, yielding the predicted value of 0.403 according to $\W(\eta)=8uv\eta^2(1-\eta)^2$.
Note that $C$-$\Gamma$ coexistence is assessed directly from the intensity moments of the mixed Gaussian source.

At the complementary settings $\eta=0.4$ and $0.6$, the measured $\W_{\rm cv}$ are nearly equal ($0.384\pm0.025$ and $0.388\pm0.031$). Both witness $C$-$\Gamma$ coexistence, but their values alone do not identify which is which.
To distinguish these two mixtures, an independent even-symmetric pseudothermal light beam produced by another SLM is introduced through a beam splitter with adjustable power.
As shown in Fig.~\ref{fig:3}(b), linear fits to the $R$-$Q$ scans yield pseudocovariance strengths $b=0.304$, $0.212$, and $0.130$ for nominal mixing fractions $\eta=0.4$, $0.5$, and $0.6$, respectively. The corresponding covariance strengths, obtained from $a=Q(0)-b$, are $0.173$, $0.249$, and $0.355$. The two channels have comparable strengths near equal mixing, while the dominant contribution switches from pseudocovariance at $\eta=0.4$ to covariance at $\eta=0.6$. Using the $C$-only endpoint calibration, the reconstructed covariance-source power fractions, $1-\sqrt{b/v}$, are $0.401\pm0.028$, $0.500\pm0.031$, and $0.608\pm0.037$, consistent with the nominal values. Refer to Secs.~S7.B--S7.C for more analysis.
Although endpoint calibration provides an additional quantitative check of the protocol, neither coexistence witnessing nor thermal-probe reconstruction of the channel strengths requires the two source contributions to be isolated and measured separately.

Having identified the channel strengths, we finally demonstrate recovering their relative phase via an equal-power coherent reference ($P=n_2$). An aperture confines the coherent reference to the region where $E_2$ is detected, leaving the centrosymmetric region where $E_1$ is measured free of reference light.
We then scan the reference phase using the SLM from $\phi=0$ to $2\pi$ with a spacing of $\pi/8$.
As shown in Fig.~\ref{fig:4}(a), the $C$-$\Gamma$ mixtures exhibit a pronounced second-harmonic modulation and reconstruct elongated ellipses, while the single-channel endpoints give nearly circular trajectories.
Figure~\ref{fig:4}(b) fixes $\eta\simeq0.5$ and adds a uniform phase $\beta$ to the binary-phase component, i.e., $\phi_C(\bm r)\in\{\beta,\pi+\beta\}$.
This multiplies $C$ by $\mathrm{e}^{2\mathrm{i}\beta}$, so the model predicts an ellipse rotation by $\beta$ at fixed channel strengths. The reconstructions follow this rotation while approximately preserving the major-axis length.
 These parameters remain fixed across both scans. Since $u\ne v$, equal source fractions need not produce an exact line. The phase-scan analysis and endpoint calibration are given in Secs.~S7.C--S7.D.

\subsection*{Spatially resolved correlation-ellipse textures}
\begin{figure*}[t]
 \centering
 \includegraphics[width=0.94\textwidth]{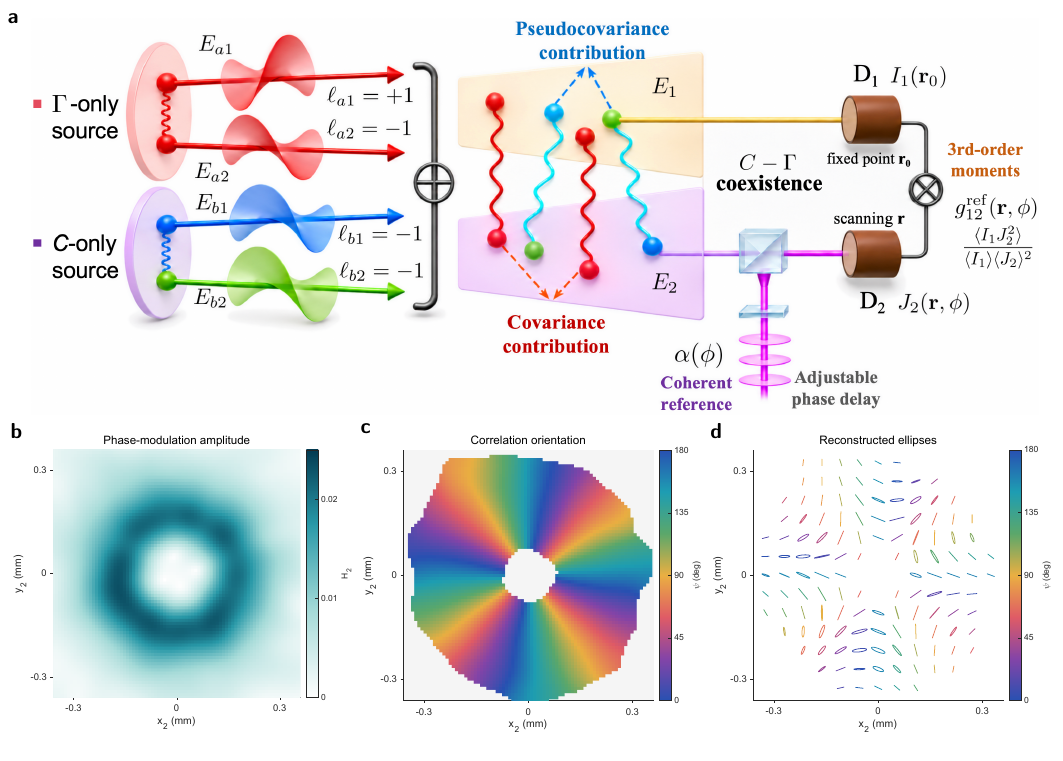}
 \caption{Spatially resolved correlation-ellipse tomography. (a) Superposition of independent $\Gamma$-only and $C$-only source pairs as in Eq.~\eqref{eq:em_sourceoutputs}, with $(\ell_{a1},\ell_{a2})=(+1,-1)$ and $(\ell_{b1},\ell_{b2})=(-1,-1)$. A coherent reference $\alpha(\phi)$ displaces arm~2 alone. Detector $D_1$ records $I_1(\bm r_0)$ at a fixed point, and $D_2$ records $J_2(\bm r,\phi)$ at the scanned point. The paired intensities give $g_{12}^{\rm ref}(\bm r,\phi)=\av{I_1 J_2^2}/(\av{I_1}\av{J_2}^2)$. Panels (b--d) set $\bm r_0=\bm0$. (b) Second-harmonic amplitude $H_2$ extracted from eight phase steps. (c) Correlation-ellipse orientation $\psi$ modulo $180^\circ$. (d) Reconstructed ellipses; color encodes $\psi$ and glyph semiaxes use a common display scale. Panels (c,d) retain $H_2>0.2H_{2,\max}$. The axes in (d) use Gaussian-constrained fits, including the line limit $B=0$; traversal arrows are omitted because the phase scans alone do not label the channels. Numerical results use $10^6$ realizations, $\eta=1/2$, and $\beta=0$.}
 \label{fig:em_spatial_source}
\end{figure*}

We next use numerical simulations to extend the measurement hierarchy from a fixed mode pair to a spatial texture of correlation ellipses. Figure~\ref{fig:em_spatial_source}(a) combines two statistically independent zero-mean Gaussian source pairs, $(E_{a1},E_{a2})$ and $(E_{b1},E_{b2})$. At source-plane position $\bm\rho=(\rho,\varphi_\rho)$, their spiral-modulated outputs are
\begin{equation}
 \begin{aligned}
 E_1&=E_{a1}\mathrm{e}^{\mathrm{i}\ell_{a1}\varphi_\rho}
       +E_{b1}\mathrm{e}^{\mathrm{i}\ell_{b1}\varphi_\rho},\\
 E_2&=E_{a2}\mathrm{e}^{\mathrm{i}\ell_{a2}\varphi_\rho}
       +E_{b2}\mathrm{e}^{\mathrm{i}\ell_{b2}\varphi_\rho}.
 \end{aligned}
 \label{eq:em_sourceoutputs}
\end{equation}
All envelopes in this equation are evaluated at $\bm\rho$; $E_{\mu i}$ denotes source $\mu$ in arm $i$ before spiral modulation and includes its weight and radial pupil. The allowed source-pair cross moments are the covariance $\Gamma_a$ and pseudocovariance $C_b$; $C_a=\Gamma_b=0$ at all pairs of source positions. The pairs are independent of each other, but their members are correlated. Proper marginal random fields ensure circular Gaussian outputs after linear propagation. Detector~1 samples a fixed far-field point, while detector~2 scans with a calibrated coherent reference added only to $E_2$.

For the matched source realization in Sec.~S9 of the Supplemental Material, with a real radial pupil and equal spatial-white kernels, the covariance carries the charge difference $m_\Gamma=\ell_{a1}-\ell_{a2}=+2$, whereas the pseudocovariance carries the charge sum $m_C=\ell_{b1}+\ell_{b2}=-2$. With detector~1 at the far-field origin, Fourier propagation gives
\begin{equation}
 \begin{aligned}
 \gamma(\bm r)&=\eta h_2(q_\perp)\mathrm{e}^{2\mathrm{i}\varphi},\\
 \chi(\bm r)&=(1-\eta)h_2(q_\perp)\mathrm{e}^{-2\mathrm{i}\varphi+2\mathrm{i}\beta}.
 \end{aligned}
 \label{eq:em_spatialmoments}
\end{equation}
Here $q_\perp=kr/f$, with $k=2\pi/\lambda_0$, is the radial spatial frequency, $\eta$ is the detected covariance-source power fraction, $h_2$ the normalized radial correlation amplitude, and $\beta$ a common phase included in both $b$-source envelopes. The far-field azimuth $\varphi$ differs from the source-plane angle $\varphi_\rho$ and the reference phase $\phi$ below. The local ellipse has
\begin{equation}
 \begin{aligned}
 A=|h_2|,\,\, B=|2\eta-1||h_2|,\,\,
 \psi=\beta-2\varphi\,\, (\rm{mod} \pi).
 \end{aligned}
 \label{eq:em_spatialellipse}
\end{equation}
Equal weights yield line trajectories; unequal weights yield ellipses with a twice-winding orientation. Their geometry corresponds to a second-order vector-vortex polarization texture~\cite{Milione2011}, with the spatial structure here residing in joint scalar-field statistics.

At each position, the displaced detector records $J_2(\bm r,\phi)=|E_2(\bm r)+\sqrt{P(\bm r)}\mathrm{e}^{\mathrm{i}\phi}|^2$. With a flat or phase-calibrated reference wavefront and $\rho(\bm r)=P(\bm r)/n_2(\bm r)$,
\begin{equation}
 \begin{aligned}
 g_{12}^{\rm ref}(\bm r,\phi)
 &=\frac{\av{I_1(\bm0)J_2(\bm r,\phi)^2}}
 {n_1(\bm0)[n_2(\bm r)+P(\bm r)]^2}\\
 &=B_0(\bm r)+\frac{4\rho(\bm r)}{[1+\rho(\bm r)]^2}
 \operatorname{Re}[\mathrm{e}^{-2\mathrm{i}\phi}\gamma^*\chi].
 \end{aligned}
 \label{eq:em_spatialreadout}
\end{equation}
Here $B_0=2-\rho^2/(1+\rho)^2+4p/(1+\rho)$. Phase cycling recovers $Z=\gamma^*\chi$ locally, and the background supplies $p=a+b$. Together, these determine the ellipse axes and orientation, and thermal channel labeling assigns its traversal sense.

For equal source weights ($\eta=1/2$, $\beta=0$), we analyze $10^6$ simulated realizations on a $300\times300$ grid at eight reference phases, $\phi_j=j\pi/4$ ($j=0,\ldots,7$). The wavelength is $632.8\,\mathrm{nm}$, the focal length $f=0.35\,\mathrm m$, and the sampled transverse pitch $9\,\mu\mathrm m$. The reference ratio $\rho(\bm r)$ is obtained from the stored reference power and local mean signal intensity. Fits retaining both first and second harmonics isolate the complex second-harmonic coefficient and hence $Z$.

The modulation amplitude $H_2=4\rho|Z|/(1+\rho)^2$ forms an annulus of radius approximately $0.17\,\mathrm{mm}$ [Fig.~\ref{fig:em_spatial_source}(b)]. Its phase yields the twice-winding orientation texture in (c). The propagation routine retains a common Fresnel phase, giving the reference-relative prediction $\psi=(2kf-\pi/2)-2\varphi\pmod\pi$. Within $H_2>0.2H_{2,\max}$, the angular root-mean-square difference from this prediction is $3.1^\circ$, without fitting an angular offset. Gaussian-constrained fits give the corresponding near-linear ellipse texture [(d)]; the raw minor-semiaxis squares fluctuate about the line limit, so the displayed axes are fitted estimates. These maps demonstrate how a single-arm phase reference reveals a spatial correlation geometry beyond the unperturbed intensity-correlation strength.

\subsection*{Discussion and Conclusion}
We experimentally prepare a programmable Gaussian source and reveal its hidden correlation structure, validating the measurement hierarchy of intensity-only $C$-$\Gamma$ coexistence witnessing through $\W_{\rm cv}>0$, thermal-probe channel identification, and single-reference correlation-ellipse tomography via phase scanning of $g_{12}(\phi)$.
A future experiment with the optical in-phase/quadrature modulators proposed in Sec.~S8 could test all three stages more precisely by comparing the measured witness, channel strengths, and ellipse orientation with values prescribed by the electrical drives.
Our results distinguish the statistical completeness of a Gaussian description from the experimental identifiability of its components. Although second-order field moments fully specify the Gaussian statistics, the intensity-correlation order needed to resolve them depends on the information sought and the available probing resources. Particularly striking is the role of added thermal fluctuations: a controlled random probe reveals information concealed in the original intensity statistics. This suggests a broader strategy for statistical optics, in which selective perturbations of field moments distinguish correlation structures that would otherwise remain observationally equivalent.

It is also implied that the statistical simplicity of Gaussian correlation can conceal a rich spatial correlation geometry.
The simulations in Fig.~\ref{fig:em_spatial_source} illustrate this possibility through correlation-ellipse textures analogous to vector-vortex polarization patterns~\cite{Milione2011}. Their local orientation is encoded in the relative phase of $C$ and $\Gamma$, which does not enter the conventional second-order intensity-correlation signal. Spatially resolved tomography thus brings the geometric language of polarization to the joint statistics of two scalar fields, opening a route to designing and retrieving optical information through spatially structured correlations.

The Supplemental Material derives the bounds on the channel strengths, the correlation ellipse, and the measurement hierarchy (Secs.~S1--S5). It treats detection of unresolved modes (Sec.~S6), documents the experimental procedures and uncertainties (Sec.~S7), and proposes an implementation using optical in-phase/quadrature modulators (Sec.~S8). Section~S9 gives the spatial model and numerical reconstruction for Fig.~\ref{fig:em_spatial_source}, including the correspondence with polarization textures. Sections~S10 and S11 derive a second-order readout with two coherent references and compare the information recovered by each measurement scheme.

\subsection*{Acknowledgments}
We wish to acknowledge the support of the National Natural Science Foundation of China (12504411, 12274037), the Basic and Applied Basic Research Foundation of Guangdong Province (2026A1515011704), and the Science and Technology Development Fund from Macau S.A.R. (FDCT) (0105/2023/RIA2); the research outputs of this Open Research Project Programme are funded by the Science and Technology Development Fund of the Macao S.A.R. (File No.: 0002/2024/TFP), Quantum Science Center of Guangdong-Hong Kong-Macau Great Bay Area (GDZX2204004, GDZX2304001), and the Fundamental Research Funds for the Central Universities.
\subsection*{Conflict of Interest}
The authors have no conflicts to disclose.
\subsection*{Data Availability}
The data and code that support the findings of this study are available from the corresponding authors upon reasonable request.

\bibliography{references}

\end{document}